\documentclass[preprint,showkeys,aps,prb]{revtex4}
\usepackage[dvips]{graphicx}
\begin{document}

\title{
Compressible Baker Maps and Their Inverses .    \\
A Memoir for Francis Hayin Ree [ 1936-2020 ]    \\
}

\author{
William Graham Hoover                           \\
Ruby Valley Research Institute                  \\
Highway Contract 60, Box 601                    \\
Ruby Valley, Nevada 89833                       \\
}

\date{\today}


\vspace{0.1cm}

\begin{abstract}
This memoir is dedicated to the late Francis Hayin Ree, a formative influence shaping
my work in statistical mechanics.  Between 1963 and 1968 we collaborated on nine papers
published in the Journal of Chemical Physics. Those dealt with the virial series,
cell models, and computer simulation. All of them were directed toward understanding
the statistical thermodynamics of simple model systems.  Our last joint work is
also the most cited, with over 1000 citations, ``Melting Transition and Communal
Entropy for Hard Spheres'', submitted 3 May 1968 and published that October. Here 
I summarize my own most recent work on compressible time-reversible two-dimensional
maps.  These simplest of model systems are amenable to computer simulation and are
providing stimulating and surprising results.
\end{abstract}

\maketitle

\section{Francis Hayin Ree [ July 1936--January 2020 ]}

In 1962 I was hired as a physicist (despite Master's and Doctor's degrees in physical chemistry)
by the Lawrence Radiation Laboratory in Livermore, California. I had read that Berni Alder
and Tom Wainwright were developing molecular dynamics there\cite{b1}, working with hard disks and
spheres, closely related to my parallel hard-square and hard-cube thesis work under Andy De Rocco at Ann
Arbor\cite{b2,b3}. Francis Ree was already established at Livermore, on LRL's square mile. He was fresh
from his own doctoral work, at the University of Utah under Henry Eyring, on random walks\cite{b4}.
Both of us were happy to have access to Livermore's tremendous computational power, along with help
from the Laboratory's hundreds of stimulating Ph Ds. Francis and I shared the computational services
of Warren Cunningham.  Warren kindly punched the cards and fetched the printouts, by Laboratory bicycle,
as we explored the application of statistical mechanics to simple models, settling into lifelong regimens
of research and publication.

\section{Nonequilibrium Simulations {\it versus} Carbon Compounds}

From the hard-sphere equation of state and integral equations for few-body distribution functions 
Francis concocted general predictive recipes for pressure-volume-energy equations of state for air,
water, hydrocarbons, and high explosives.  His enthusiasm for phase transitions, honed on simple
statistical models and the rare gases, carried over to carbon with its graphite and diamond phases.
Francis generated accurate thermodynamic data for pentaerythritol tetranitrate, a high-explosive
relative of nitroglycerine. He described the behavior of polybutene and phase-diagram behaviors of
the rare gases, graphite and diamond, as well as many of carbon's compounds, both organic and
inorganic\cite{b5}. He became a permanent member of the Korea Academy of Science and Technology.  He
received the Department of Energy's Award of Excellence and two Lawrence Livermore National Laboratory
Distinguished Achievement Awards after contributing thirty years' work to the Livermore Laboratory. A
scholar and a gentleman.

By 1970 Francis' and my research paths had separated.  Berni Alder had helped me to a Professorship at
the University of California's Department of Applied Science.  There I worked with my first Ph D
student, Bill Ashurst. We set out to develop nonequilibrium atomistic solution techniques, ``nonequilibrium
molecular dynamics''\cite{b6}. Meanwhile Francis was drawn to more realistic representations of the
relatively complex systems supporting the Livermore Laboratory's weapons programs.  He carried out over
100 research inquiries with dozens of coauthors. Most of his work had its base in relatively esoteric
elaborations of equilibrium statistical mechanics. That underlying theory was applied to down-to-earth
practical applications for materials relevant to weapons research.

Meanwhile my own work progressed along lines more academic than applied, facilitated by my joint
appointments in the University and the Laboratory. Bill Ashurst and I mostly restricted our collaborative
work to the Lennard-Jones pair potential, a crude representation of argon, developing simulations of its
equilibrium and nonequilibrium properties (viscosity and thermal conductivity). In inventing
nonequilibrium boundary conditions for the latter we soon discovered, in retrospect and to our surprise,
that all the nonequilibrium motion equations we had fashioned were time-reversible, so that any short
trajectory fragment, advancing from one time to another, could be precisely reversed. One could do so by
changing the signs of the momenta as well as all of the time-reversible friction coefficients
(described in the next Section) used to impose thermal boundary conditions on the simulations. The most
surprising consequence of the simulations, that irreversible behavior was generated by time-reversible
equations of motion, fascinated us just as it had Boltzmann a century earlier.  In the next Section we
explore examples of time reversibility with two models, one of them with the desirable property of
``ergodicity'' and the other one not.

\section{Nonequilibria, Time Reversibility, and Ergodicity}

\subsection{Nonequilibrium Thermostatted Systems}

Following my six years' experience working with Francis on virial series (density expansions of the
pressure), and equation of state problems I was convinced that equilibrium statistical mechanics and
corresponding computer simulations were understood sufficiently well. I set out to study nonequilibrium
systems driven by differences in velocity or temperature.  Invariably I sought out the simplest possible
systems for detailed studies.  According to kinetic theory diffusion can be studied by following a single
moving particle through an array of scatterers.  Simulating viscous flow requires at least two
oppositely-moving particles, and heat flow three, or so I thought then.

Some 30 years later, in 1997, Harald Posch and I formulated two new {\it single-particle} models for
thermostatted heat conductivity\cite{b7}. In both these cases we studied the motion of a single
nonequilibrium oscillator exposed to a smoothly-varying temperature gradient $(dT/dq)$ with a maximum
value of $\epsilon$ :
$$
T(q) \equiv 1 + \epsilon \tanh(q) \rightarrow \{ \ 1-\epsilon < T < 1+\epsilon \ ; \
| \ (dT/dq) \ | \le \epsilon \ \} \ .
$$
This coordinate-dependent temperature $T(q)$ was controlled with either a single Nos\'e-Hoover\cite{b8}
``thermostat'' variable $\zeta$ :
$$
\{ \ \dot q = p \ ; \ \dot p = -q - \zeta p \ ; \ \dot \zeta = p^2 - T(q) \ \} \ [ \ \rm{NH} \ ] \ ,
$$
or, following my 1996 work with Brad Holian\cite{b9}, with two such control variables, $\zeta$ and $\xi$ :
$$
\{ \ \dot q = p \ ; \ \dot p = -q - \zeta p - \xi p^3 \ ; \ \dot \zeta = p^2 - T(q) \ ; \
\dot \xi = p^4 - 3p^2 \ \} \ [ \ \rm{HH=PH} \ ] \ .
$$
Here the time-reversible controls, $\zeta$ and $\xi$, provided the initial conditions allow it, drive a
purely-kinetic heat current $(p^3/2)$ and generate a nonequilibrium steady state in three-dimensional
$(q,p,\zeta)$ or four-dimensional $(q,p,\zeta,\xi)$ phase space. The steady state is the time-averaged
probability density in the phase space. In the ``equilibrium'' case, where temperature is constant, $T=1$,
the time-averaged second and fourth moments are constrained by $\zeta$ and $\xi$ to the values from Gibbs'
canonical ensemble with $T \equiv 1$, $\langle \ p^2,p^4 \ \rangle = 1,3$ .

The three- and four-dimensional descriptions of a nonequilibrium oscillator turned out quite differently
to the predictions of Gibbs' ensembles. Nonequilibrium distributions are typically ``fractal'' and can be
quite intricate far from equilibrium. {\bf Figure 1} compares a two-dimensional $(q,p)$ phase-space cross
section cut through a four-dimensional representation of HH=PH dynamics to the corresponding cross section
using HH=SHH dynamics.

\begin{figure}
\includegraphics[width=2.0 in,angle=-90.]{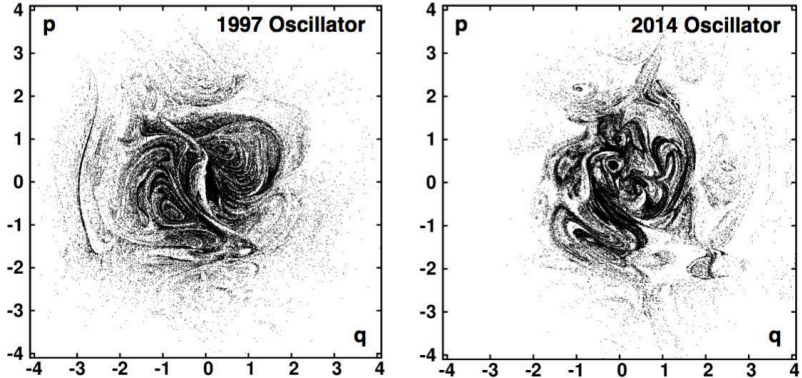}
\caption{
           $p(q)$ sections with $\zeta = \xi = 0$ for two versions of a conducting oscillator
           with both quadratic and quartic moments controlled by the PH and SHH generalizations
           of the Hoover-Holian motion equations given in the text. Here the maximum temperature
           gradient $\epsilon$ is 0.40.
}
\end{figure}

In both cases the imposed temperature varies in space, $T(q) \equiv 1 + 0.4\tanh(q)$. In my more recent 2014
work with Clint Sprott\cite{b10} and my Wife Carol we used a more-elaborate motion equation for the quartic
control variable $\xi$ :
$$
\{ \ \dot q = p \ ; \ \dot p = -q - \zeta p - \xi p^3 \ ; \ \dot \zeta = p^2 - T(q) \ ; \
\dot \xi = p^4 - 3p^2T(q) \ \} \ [ \ \rm{HH=SHH} \ ] \ .
$$
The comparison of the two fractal cross sections given in {\bf Figure 1} shows that the two similar
$\dot \xi$ control equations produce very different fractals far from equilibrium.

\subsection{Time Reversibility, Dissipation, and the Second Law of Thermodynamics}

Notice that both the Nos\'e-Hoover and the Hoover-Holian sets of motion equations {\it are} ``time-reversible''.
By this I mean that changing the signs of the time, the momentum $p$, and the friction coefficients $\zeta$ and
$\xi$ precisely reverses the time-development of the coordinate $q$.  An undesirable feature of the simpler
Nos\'e-Hoover thermostat model is that it often lacks ergodicity\cite{b8,b10} when applied to small systems.
In the conducting oscillator example problem the three-dimensional $(q,p,\zeta)$ phase space contains
infinitely-many stationary states, mostly two-dimensional tori with no nonequilibrium heat flux. This
disturbing abundance of unphysical solutions can be cured by introducing the additional control variable $\xi$
which originated in the Hoover-Holian example\cite{b9}.

With $\xi$ included, the resulting phase-space distributions are typically fractal (fractional-dimensional)
attractors, as suggested by the two sample cross-sections of {\bf Figure 1}.  In such fractals the distribution
of time-reversible $(q,p,\zeta,\xi)$ states describes an irreversible dissipative flow of kinetic energy
{\it from} a mirror-image repellor {\it to} its attractor.  The resulting hot-to-cold direction of the mean
energy flow is just that mandated by the macroscopic Second Law of Thermodynamics.  Though the additional control
variable $\xi$ increases the dimensionality of phase space from three to four this extra complexity seems a
reasonable price to pay for the simplicity of ergodicity.

\subsection{Ergodicity and Ergodic Nonequilibrium Fractals}

\begin{figure}
\includegraphics[width=2.0 in,angle=-90.]{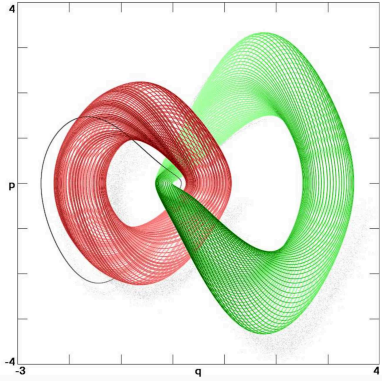}
\caption{
           Three disconnected solutions for the Nos\'e-Hoover oscillator with $\epsilon = 0.42$.  The red
           and green toroidal solutions and the black dissipative limit cycle all satisfy exactly
           the same nonequilibrium motion equations, with $T(q) = 1 + \epsilon \tanh (q)$.
}
\end{figure}

At equilibrium ergodicity is crucial for the validity of Willard Gibbs' statistical mechanics.  Gibbs was
able to formulate macroscopic thermodynamic properties as microscopic phase-space averages provided that
all the phase-space states included in the averages were accessible dynamically, one from another.
An ``ergodic'' system is one in which the phase-space average is equivalent to a longtime dynamical
average.  Evidently the thermostatted Nos\'e-Hoover oscillator just considered is not ergodic --
each of the three solutions of {\bf Figure 2} is confined to its own one- or two-dimensional portion
of the three-dimensional $(q,p,\zeta)$ space.  In striking contrast the two similar Hoover-Holian oscillators
of {\bf Figure 1} used two friction coefficients and provided ergodic nonequilibrium steady states independent
of initial conditions\cite{b8,b9,b10}.

The fractal nature of nonequilibrium flows is typical of ergodic stationary states\cite{b11,b12}. Such 
fractals are invariably chaotic and anisotropic, with an overall positive Lyapunov exponent describing the mean
(time-averaged) separation rate of two nearby phase-space trajectories.  Additionally, one or more negative
exponents cause the overall negative sum responsible for the zero-phase-volume steady-state structures of
fractal phase-space attractors: $d\ln(\otimes)/dt \equiv\sum \lambda_i < 0$.  Here $\otimes$ is an infinitesimal
element of comoving phase-space volume.  Because the summed-up Lyapunov exponents on the repellor (with
their opposite signs) are positive rather than negative any nearby trajectory undergoes an exponentially-fast
departure from the well-named repellor to the dissipative strange attractor.  The typical case, in which both
structures have a presence throughout the phase space, renders the mental picture of repellor-to-attractor
flow somewhat paradoxical.

By 1987 it became clear that phase-space distributions for nonequilibrium stationary states were typically
fractal and ergodic in character\cite{b11,b12} with a wide assortment of fractional dimensions which were
defined and described by R\'enyi, Mandelbrot, and a huge literature of follow-on work. Harald Posch and I
found that two-dimensional maps, clearly simpler than three- and four-dimensional flows, could be chosen to
model the properties of time-reversible nonequilibrium simulations\cite{b13}. We chose to study the
compressible time-reversible Baker Map shown at the left of {\bf Figure 3}.  Iterating the  map converts one
$(q,p)$ state to another and another and another \dots, rather like the sequence of discrete movie frames
used to describe a continuous motion. Extensions of this work\cite{b14}, some very recent\cite{b15,b16}, have
confronted us with some surprising and stimulating results. I describe those next.

\begin{figure}
\includegraphics[width=1.5 in,angle=-90.]{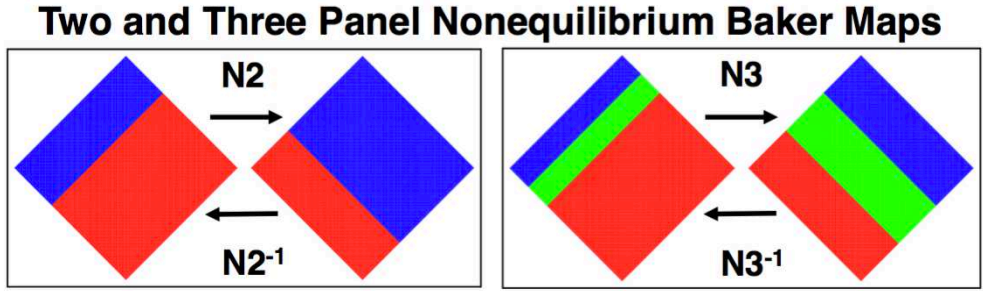}
\caption{
           Tricolor representations of the N2 and N3 maps. In both cases the rectangular areas prior
           to mapping are deformed without rotation, shrinking and expanding in orthogonal directions.
           Iterations of these linear maps and their inverses provide fractal structures resembling
           those generated with nonlinear dissipative flows. See Figure 4.
}
\end{figure}

\section{Maps Caricaturing Nonequilibrium Statistical Flows}

Time-reversible molecular dynamics can be applied equally well to equilibrium systems
and to nonequilibrium atomistic flows of mass, momentum, and energy. Despite reversibility
the motion equations used to model steady flows invariably produce dissipative
irreversibility, as is required by the Second Law of Thermodynamics\cite{b11,b12}. One
can best seek to grasp an understanding of how reversible equations provide irreversible
behavior by the study of small systems well suited to support using computer graphics.
Such investigations revealed that the mechanism for computational irreversibility is a
consequence of the fractal repellor-to-attractor nature of time-reversible nonequilibrium
phase-space states.

I turn here from flows to maps, stressing recent results. This choice simplifies and clarifies
analyses. Like flows, maps are deterministic, can be time-reversible, and often generate fractals.
Their phase-space compressibility rate, $(d\ln \otimes/dt) = \sum \lambda_i$, is closely related
to the Gibbs-entropy production production rate $\dot S={\rm k}\ln\dot\otimes$ and to the fractal
information dimension, just as is the case with flows.  Here $\otimes$ is an infinitesimal element
of phase volume comoving with the flow and k is Boltzmann's constant.

There is one interesting qualitative difference between flows and maps: irreversibility in flows
stems from nonlinearity while even the simplest linear maps can illustrate the dissipative behavior
obeying the Second Law. Let us next compare the compressible but time-reversible linear Baker Map
of 30 years ago to a similar one which lacks time reversibility but retains the dissipative fractal
character of typical nonequilibrium flows. 

\section{Of Two Linear Baker Maps, N2 is Reversible While N3 is Not}

\begin{figure}
\includegraphics[width=2. in,angle=0.]{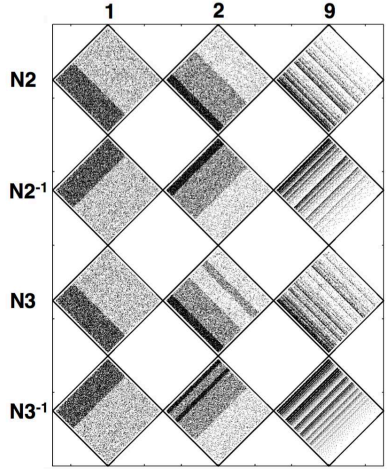}
\caption{
           Results obtained by iterating N2 and N3 and their inverses once, twice, and nine times.
           Notice that the iterations for N2 forward and backward are mirror images due to the
           time-reversibility of those maps. The lack of this symmetry for iterations of N3 and
           N3$^{-1}$ shows that those more complex maps lack time-reversibility.
}
\end{figure}

\begin{figure}
\includegraphics[width=3. in,angle=-90.]{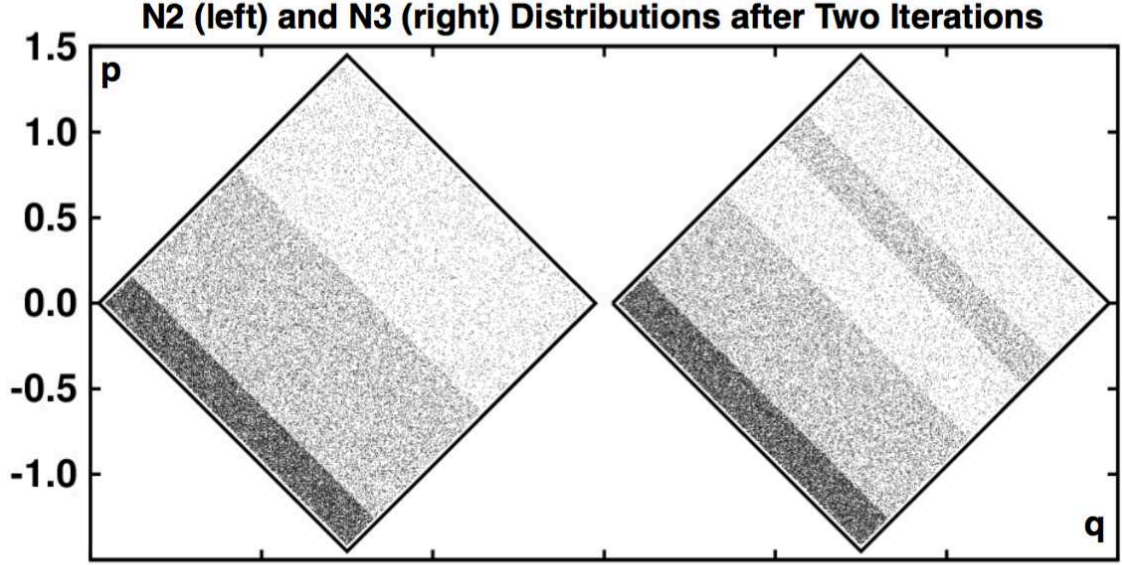}
\caption{
           The distributions of 50,000 random initial points after two forward mappings of N2 and
           N3. The three different densities of points, relative to a unit square,  are 4, 1, and
           1/4, giving identical information dimensions for the two distributions shown here.
}
\end{figure}

Two-Dimensional Maps are analogous to cross-sections of Three-Dimensional Flows.  Here 
we consider the interesting fractal structures generated by two similar simple linear maps,
N2 and N3, which operate within the finite diamond-shaped domains shown in {\bf Figures 3
and 4}. The reason for the diamond-shaped $(q,p)$ domains chosen here is linked to time
reversibility. Time Reversibility is a stringent constraint on maps.  It requires that the
``before'' and ``after'' regions have identical shapes and sizes. This requirement is easily
satisfied with N2 where there is a single discontinuity in the mapping between two rectangles
with different areas. Here time reversibility is easy to check.  We adopt the conventional
three-step meaning: the effect of the mapping can be reversed to reach the previous
condition, by carrying out three successive steps, T*N2*T.  These correspond to [1] time-reversal,
[2] forward mapping, and [3] a second time reversal. Here T is the time-reversal map
[T$(\pm q,\pm p)=(\pm q,\mp p)$], changing the sign of the momentum $p$ but leaving the coordinate
$q$ unchanged.

Baker Maps take a point in two-dimensional $(q,p)$ [ coordinate, momentum ] space
from one iteration to the next.  Such a mapping can be applied to individual points
or to areas in two-dimensional $(q,p)$ phase space.  Two similar linear Baker Maps,
along with their inverses confined to the same portion of two-dimensional space, are
illustrated in {\bf Figures 3-7}. I have written the analytic forms of all the maps so
that they occupy diamond-shaped domains with maximum and minimum values of the horizontal
coordinate $q$ and the vertical momentum $p$ equal to $\pm \sqrt{2}$ so that the map area
is 4 and the diamonds' centers are at the origin.

In describing the density within these diamonds it is convenient to imagine them mapped to
a unit square with the density integrated over the square equal to unity.  In speaking of the
density in discussing Figures 5 and 9 and 10 we will adopt the picture that the probability 
density occupies a unit square and that the total density integrates to unity.

In the $(q,p)$ coordinate system appropriate to diamond domains the analytic forms of the
simplest time-reversible N2 map and its inverse are as follows :

\begin{verbatim}
if(q-p.lt.-4*d) qnew = + (11/ 6)*q - ( 7/ 6)*p + 14*d
if(q-p.lt.-4*d) pnew = - ( 7/ 6)*q + (11/ 6)*p - 10*d
if(q-p.ge.-4*d) qnew = + (11/12)*q - ( 7/12)*p -  7*d
if(q-p.ge.-4*d) pnew = - ( 7/12)*q + (11/12)*p -  1*d
[ Reversible Nonequilibrium Baker Map N2 with d = sqrt(1/72) ]
\end{verbatim}
The inverse mapping N2$^{-1}$ follows easily from the linear equations, with the result :
\begin{verbatim}
if(q+p.lt.-4*d) qnew = + (11/ 6)*q + ( 7/ 6)*p + 14*d
if(q+p.lt.-4*d) pnew = + ( 7/ 6)*q + (11/ 6)*p + 10*d
if(q+p.ge.-4*d) qnew = + (11/12)*q + ( 7/12)*p -  7*d
if(q+p.ge.-4*d) pnew = + ( 7/12)*q + (11/12)*p +  1*d
[Inverse of the Nonequilibrium Baker Map N2 ]
\end{verbatim}

The time-reversibility of the N2 map guarantees that the inverse mapping is equivalent
to the three-step process mentioned above, T*N2*T.  For flows this reversal analog for a
timestep $+dt$ corresponds to taking a {\it negative} timestep $-dt$ and then changing the
sign of the momentum.  Because all the maps we consider here are Lyapunov unstable (more
about this later) the practical length of a reversed trajectory is limited by the exponential
growth of roundoff error.  With quadruple-precision arithmetic a typical sequence of N2 or N3
mappings can be recognizably reversed for about fifty iterations.

The N2 and N3 mappings of {\bf Figures 3-7} are similar, but differ in one fundamental way.
The N3 mapping is {\it not reversible}, though Carol and I  mistakenly thought that it
was\cite{b15}.  The inverse mapping follows easily from the linear equations just as in the N2
case. But  the inverse is not at all the same as the three-step mapping T*N*T which applies in
the time-reversible case.

\pagebreak

\begin{verbatim}
if (q-p.lt.-8*d)                    qnew = +19*q/ 6 - 17*p/ 6 + 34*d
if (q-p.lt.-8*d)                    pnew = -17*q/ 6 + 19*p/ 6 - 26*d
if((q-p.ge.-8*d).and.(q-p.le.-4*d)) qnew = +19*q/ 6 - 17*p/ 6 + 18*d
if((q-p.ge.-8*d).and.(q-p.le.-4*d)) pnew = -17*q/ 6 + 19*p/ 6 - 18*d
if (q-p.gt.-4*d)                    qnew = +11*q/12 -  7*p/12 -  7*d
if (q-p.gt.-4*d)                    pnew = - 7*q/12 + 11*p/12 -  1*d
[ Irreversible Nonequilibrium Baker Map N3 with d = sqrt(1/72) ]
\end{verbatim}
\begin{verbatim}
if (q+p.ge.+4*d)                    qnew = +19*q/12 + 17*p/12 - 17*d
if (q+p.ge.+4*d)                    pnew = +17*q/12 + 19*p/12 -  7*d
if((q+p.lt.+4*d).and.(q+p.gt.-4*d)) qnew = +19*q/12 + 17*p/12 -  3*d
if((q+p.lt.+4*d).and.(q+p.gt.-4*d)) pnew = +17*q/12 + 19*p/12 +  3*d
if (q+p.le.-4*d)                    qnew = +11*q/ 6 +  7*p/ 6 + 14*d
if (q+p.le.-4*d)                    pnew = + 7*q/ 6 + 11*p/ 6 + 10*d
[ Inverse of the Nonequilibrium Baker Map N3 ]
\end{verbatim}

{\bf Figures 4-7} show the evolution of the fractals, comparing the forward and
backward developments for N2, N3, and their inverses.  {\bf Figure 4} compares the
distributions after one, two, and nine forward mappings. The detailed structures of the
limiting attractors for all four maps are shown in {\bf Figures 6 and 7}. Notice
particularly that the N3 attractor and repellor are quite different fractals, not
mirror images of one another and with different fractal dimensions.  This is confirmation
that N3 and its inverse are not time-reversible.

\begin{figure}
\includegraphics[width=3. in,angle=-90.]{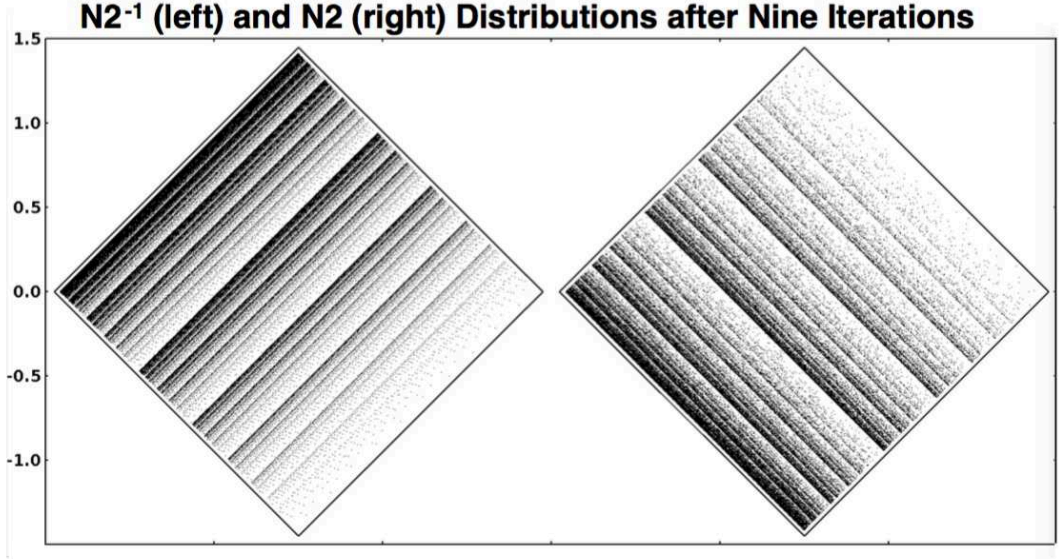}
\caption{
           Repellor (left) and Attractor (right) for N2, corresponding to nine
           iterations, visually quite similar to the limiting infinite-iterations
           case. 200,000 points are shown.  The repellor/attractor pair are mirror
           images of each other where the mirror is horizontal.
}
\end{figure}

\begin{figure}
\includegraphics[width=3. in,angle=-90.]{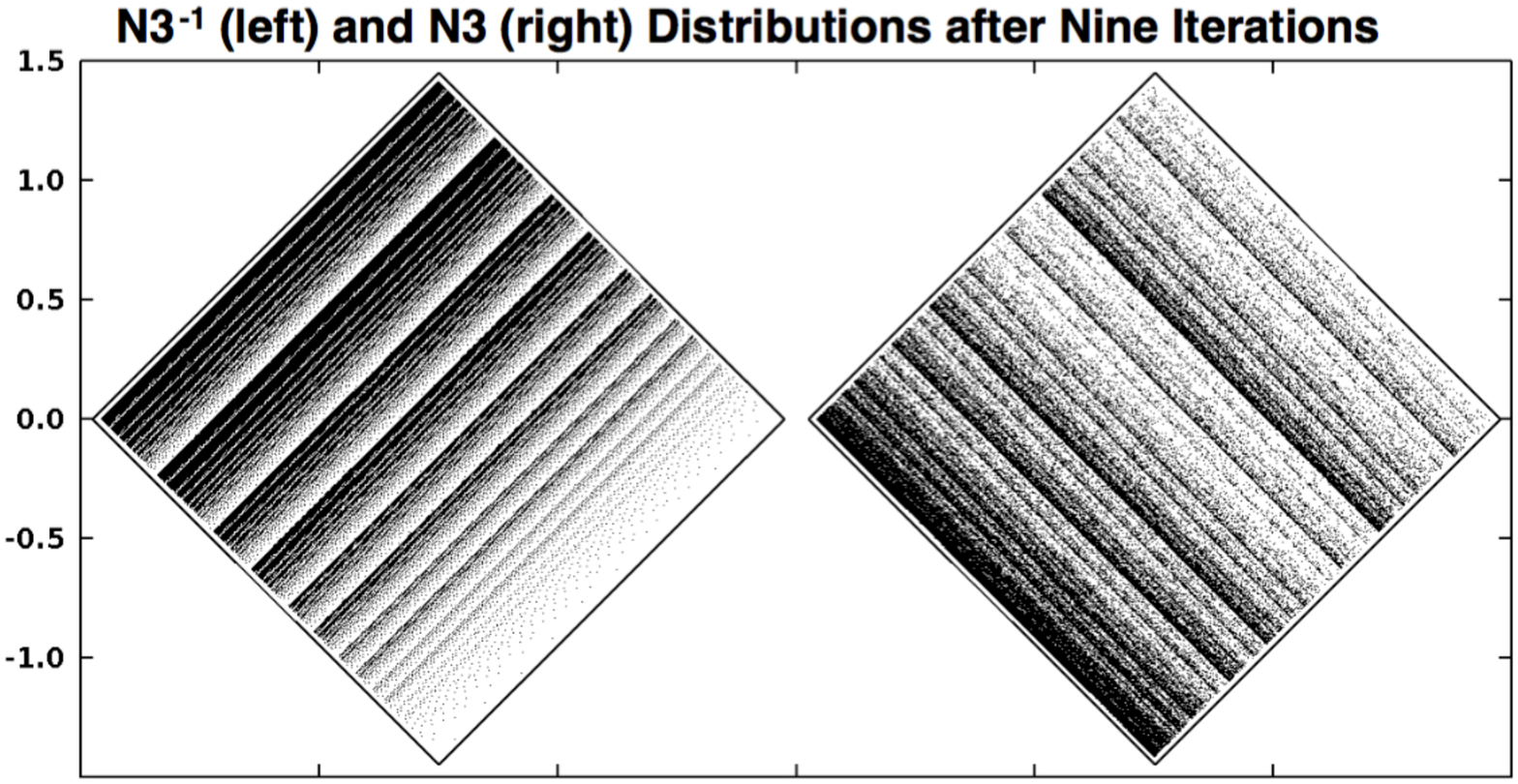}
\caption{
           Repellor (left) and Attractor (right) for N3.  Notice that time-reversibility
           symmetry is absent.  The repellor generated by N3$^{-1}$ is also fractal, but
           with a different information dimension than that of the attractor, and with a
           much steeper falloff of density, as shown in Figure 9.
}
\end{figure}

\section{The Irreversible N3 Map and Its Information Dimension}

The N3 map was motivated by the idea to symmetrize the ``after'' image of the N2 map
shown in {\bf Figure 3} to provide the three equal-area regions shown at the right of
the Figure.  Although the linear equations describing the N3 map and its inverse are
readily constructed and given above, we can see that the attractor and repellor shown
in {\bf Figure 7} are not mirror images.  They are particularly simple from
a structural point of view. It is easy to verify that T*N3*T, where T reverses the
sign of the vertical ( momentum ) coordinate $p$, does not result in the inverse
N3$^{-1}$.

Consider the action of N3 on a uniform distribution. The first iteration
results in three strips of equal width (red, green, and blue at the right of {\bf Figure 3}),
with densities $\{2,\frac{1}{2},\frac{1}{2}\}$ and probabilities $\{\frac{2}{3},\frac{1}{6},
\frac{1}{6}\}$.  The corresponding information dimension is :
$$
\textstyle{
D_I = 1 + [\frac{2}{3}\ln \frac{2}{3} + \frac{1}{3} \ln \frac{1}{6} ]/\ln\frac{1}{3} = 1.78969 \ .
}
$$
A second iteration (third row and second column of {\bf Figure 4} provides nine strips of width
$\frac{1}{9}$, with densities $\{ 4,1,1,1,\frac{1}{4},\frac{1}{4},1,\frac{1}{4},\frac{1}{4} \}$,
giving for the information dimension :
$$
\textstyle{
D_I = 1 + [\frac{4}{9}\ln \frac{4}{9} + \frac{4}{9} \ln \frac{1}{9} +
\frac{1}{9} \ln \frac{1}{36} ]/\ln \frac{1}{9} = 1.78969 \ .
}
$$
Continuing the iteration of the map produces no change, just repetition of the
result $D_I({\rm N3}) = 1.78969$.

N2 on the other hand provides a different distribution of densities, but only
different in their ordering :
$\{4,1,1,1,1,\frac{1}{4},\frac{1}{4},\frac{1}{4},\frac{1}{4}\}$ . {\bf Figure 5} compares
the distributions according to N2 and N3 after just two iterations of the mapping, starting
with a uniform distribution.  The three different densities of points shown in the
distributions differ from each other by powers of 4, as illustrated explicitly above.
Evidently, by construction, the information dimensions for the two maps are identical,
equal to 1.78969.  In the next Section we will see that this is false!

\section{Symmetry of the Repellor and Attractor}

\begin{figure}
\includegraphics[width=3. in,angle=-90.]{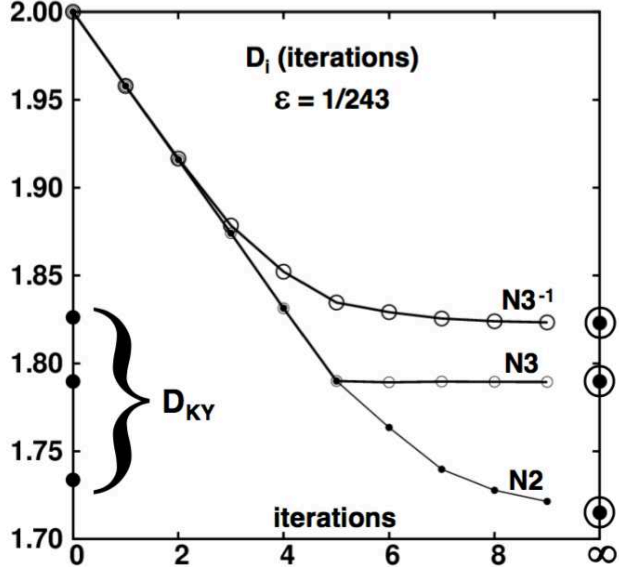}
\caption{
           Information dimensions for N2, N3, and N3$^{-1}$ using 243 mesh points as functions
           of the number of map iterations. $D_I$ remains the same for N2 and N3 through five
           iterations, but their limiting information dimensions (circled) differ. Further
           mesh refinements extend the agreement but in the end the convergence remains
           nonuniform. The Kaplan-Yorke conjectured dimensions are shown at the left and suggest
           that N3 and its inverse obey that conjecture while N2 may not.
}
\end{figure}

Although the attractors of the N2 and N3 maps appear to have the same information dimensions, as
quantified above, numerical work tells a different story, and in an interesting way, illustrated
in {\bf Figure 8}.  There we choose a particular mesh, $3^{-5} =\frac{1}{243}$, and compute the
information dimension after each of several iterations. Nine, as well as the limiting value, are
shown in the figure. Through the fifth iteration the information dimensions for N2 and N3 agree
precisely. After that iteration the dimension of the N3 map remains unchanged at
1.78969 while that of N2 continues to fall, as shown in {\bf Figure 8}.  Careful numerical work
on the N2 problem\cite{b16,b17} has suggested two different values (!) for the information dimension
of N2, neither of them equal to 1.78969: 1.7337 and 1.741$_5$.
The first of these is the Kaplan-Yorke estimate,
$$
D_{\rm KY} = 1 - (\lambda_1/\lambda_2) \ .
$$
where the Lyapunov exponents $\lambda_1 > 0$ and $\lambda_2 < 0$ are evaluated in the following Section.
The second estimate $D_I\stackrel{?}{=}1.741_5$ is the result of trillion-iteration simulations of the
N2 mapping using meshes that are integral powers of (1/3) as high as $3^{-20}$.  The resolution of this
uncertainty has been set as the 2020 Snook Prize problem\cite{b15}.

\section{Lyapunov Instability of Compressible Maps}

The Lyapunov exponents of maps, like those of flows, describe the rate at which two nearby
points separate.
The N2 map gives threefold expansion with orthogonal $\frac{3}{2}$-fold compression
one third of the time (see the blue panel at the left of {\bf Figure 3} and $\frac{3}{2}$-fold expansion
with orthogonal threefold compression two-thirds of the time (the red panel), corresponding to the
Lyapunov exponents:
$$
\textstyle{
\lambda_1 = \frac{1}{3}\ln(3)+\frac{2}{3}\ln\frac{3}{2} =
\frac{1}{3}\ln\frac{27}{4} = +0.636514 \ ; \
}
$$
$$
\textstyle{
\lambda_2 = \frac{1}{3}\ln\frac{2}{3}+\frac{2}{3}\ln\frac{1}{3} =
\frac{1}{3}\ln\frac{2}{27} = -0.867563 \ .
}
$$
The N3 map can be analyzed similarly following {\bf Figure 3}, with the results
$$
\textstyle{
\lambda_1 = \frac{1}{3}\ln(6) + \frac{2}{3}\ln\frac{3}{2} =
\frac{1}{3}\ln\frac{27}{2} = 0.867563 \ ; \
}
$$
$$
\textstyle{ 
\lambda_2 = \frac{1}{3}\ln\frac{1}{3} + \frac{2}{3}\ln\frac{1}{3}=
\frac{1}{3}\ln\frac{1}{27} = -1.098612 \ .
}
$$
These results are of interest in view of the Kaplan-Yorke conjecture that the fractal
dimension for such a map is given by $D_{KY} = 1 - (\lambda_1/\lambda_2)$, 1.73368 for
N2 and 1.78969 for N3. Thus the unchanging information dimension for the N3 map with
increasing iterations agrees precisely with the Kaplan-Yorke conjecture while that of
the somewhat simpler N2 map does not.

\section{Iteration of the Maps Gives Fractal Attractors}

\begin{figure}
\includegraphics[width=2.0 in,angle=-90.]{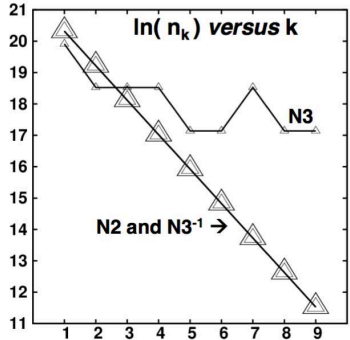}
\caption{
           Occupants of panels of width (1/3), (2/9), (4/27), \dots, or
           (1/3)(2/3)$^{k-1}$ with panel indices $1 \le k \le 9$.  Each point is taken
           from a sampling of one billion iterations.  Both N2 and N3$^{-1}$ have the
           same ``unit-square-based'' panel densities of $(1/2)^{k-2}$. The N3 mapping
           here, for nine panels of equal width, produces densities varying as powers of 4.
}
\end{figure}

The two maps give similar strange attractors.  We saw that N3 has the more ``conventional''
behavior, relative to N2, in that the information dimension after a single iteration
[applied to a constant density throughout the diamond-shaped domain] is the same as that for
many iterations and is also equal to the information dimension which follows from
the Kaplan-Yorke conjecture
$$
D_I \stackrel{?}{=} D_{KY} = 1 - (\lambda_1/\lambda_2) = 1.789690 \ .
$$
{\bf Figures 6 and 7} show the attractors and repellors for the two maps. The N2 map shows
the expected symmetry for time-reversible maps.  The repellor and attractor are mirror
images of each other.  The N3 map is different though its repellor
is ``simpler'' than the N2 fractals. The data in {\bf Figure 8} indicate that N3 and its
inverse have different information dimensions. In fact the differences are qualitative.
A look at {\bf Figure 7} correctly suggests that the fractal generated by N3 consists
of three similar panels. The density pattern in the southwest third of the figure is
repeated, with a density four times lower, in the middle and northeast thirds of the
domain.  On the other hand the attractor of the inverse map N3$^{-1}$ repeats its
northwest third (with two-thirds the density) infinitely-many times toward the southeast
in panels of widths of $(1/2)(2/3)^k$ with overall densities $2^{2-k}$ for all positive $k$.
The N2 map and its inverse behave in exactly this same way, with two thirds of their points
in the densest third of the diamond, giving densities of 2$^{2-k}$ for all positive $k$.
{\bf Figure 9} illustrates these fractal dependencies with a sampling of a billion,
e$^{20.723}$, points.  

{\bf Figure 10} shows the convergence of finite-mesh samplings of $10^{11}$ iterations,
binned into meshes as small as $(1/2)^{25}$, $(1/3)^{18}$, and $(1/6)^{11}$.  These
data indicate that both N3 and its inverse N3$^{-1}$ satisfy the Kaplan-Yorke conjecture
where the Kaplan-Yorke values are shown as the zero-mesh limits. Meanwhile the data for
N2 are confusing.  An explanation or clarification of this confusion is the subject of
the 2020 Snook Prize\cite{b18}.

\begin{figure}
\includegraphics[width=1.8 in,angle=-90.]{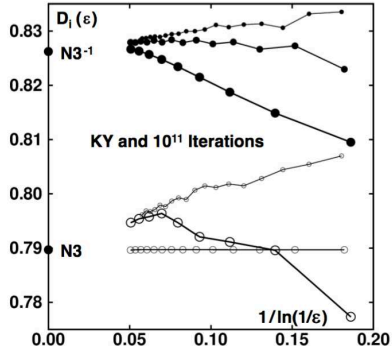}
\caption{
           Mesh-dependent information dimension estimates for the N3 and N3$^{-1}$ map
           with meshes varying from $(1/2)^{8}$ to $(1/2)^{25}$ (fine lines), $(1/3)^{5}$ to
           $(1/3)^{18}$ (medium lines), and $(1/6)^{3}$ to $(1/6)^{11}$ (thick lines).  The
           rapid convergence of the N3 map data with meshes of $(1/3)^n$  is promoted by its
           threefold symmetry, as shown in Figs. 3 and 7.
}
\end{figure}
\section{Summary}

The small-system explorations begun with Francis Ree, emphasizing simple models well-suited
to computer simulation, have provided the details of the melting and freezing transitions
for hard particles as well as the connection of thermostatted nonequilibrium simulations
to the Second Law of Thermodynamics and Irreversibility.  The finding of fractal
phase-space structures led to the investigation of maps, whose history goes back to Hopf's
work in 1937. The maps, though simpler than flows, have very recently led to findings
that are a surprise, suggesting that there is more to learn about the fractal structures
that play an important role in statistical mechanics.

\pagebreak

\section {Joint Publications of Francis Ree and the author}

We published nine joint works, all of them in the Journal of Chemical Physics, including one
written jointly with Berni Alder.  The volume numbers are indicated in this list:
Fifth and Sixth Virial Coefficients for Hard Spheres and Hard Disks {\bf 40};
On the Signs of the Hard Sphere Virial Coefficients  {\bf 40};
Reformulation of the Virial Series for Classical Fluids {\bf 41};
Dependence of Lattice Gas Properties on Mesh Size {\bf 41} [ with Berni Alder ];
Calculation of Virial Coefficients. Squares and Cubes with Attractive Forces {\bf 43};
Thermodynamic Properties of a Simple Hard-Core System {\bf 45};
Seventh Virial Coefficients for Hard Spheres and Hard Disks {\bf 46};
Use of Computer Experiments to Locate the Melting Transition and Calculate the Entropy
in the Solid Phase {\bf 47};
Melting Transition and Communal Entropy for Hard Spheres {\bf 49}.

\end{document}